\documentclass[10pt,aps,twocolumn,superscriptaddress,preprintnumbers,nofootinbib]{revtex4-1} 
\usepackage{amsmath,amssymb,graphicx,natbib,bm}
\usepackage{enumerate}
\usepackage{hyperref}
\usepackage{color}

\usepackage{empheq}
\usepackage{slashed}

\newcommand{\beq}{\begin {equation}}  
\newcommand{\eeq}{\end   {equation}} 
\newcommand{\bea}{\begin {eqnarray}} 
\newcommand{\eea}{\end   {eqnarray}}  
\newcommand{\baa}{\begin {array}   } 
\newcommand{\eaa}{\end   {array}   }     
\newcommand{\bit}{\begin {itemize} }
\newcommand{\eit}{\end   {itemize} }
\newcommand{\be }{\begin {equation}} 
\newcommand{\ee }{\end   {equation}}
\newcommand{\nn }{\nonumber        }

\begin{document}


\preprint{ACFI-T16-02}


\title{Hidden Gauged $U(1)$ Model: Unifying Scotogenic Neutrino and Flavor Dark Matter}

\author{Jiang-Hao Yu}
\email{jhyu@physics.umass.edu}
\affiliation{Amherst Center for Fundamental Interactions, Department of Physics, University of Massachusetts-Amherst, Amherst, MA 01003, U.S.A.}


\begin{abstract}
In both scotogenic neutrino and flavor dark matter models, the dark sector communicates with the standard model fermions via Yukawa portal couplings. 
We propose an economic scenario that scotogenic neutrino and flavored mediator share the same inert Higgs doublet and all are charged under a hidden gauged $U(1)$ symmetry. 
The dark $Z_2$ symmetry in dark sector is regarded as the remnant of this hidden $U(1)$ symmetry breaking.
In particular, we investigate a dark $U(1)_D$ (and also $U(1)_{B-L}$) model which unifies scotogenic neutrino and top-flavored mediator. 
Thus dark tops and dark neutrinos are the standard model fermion partners, and the dark matter could be inert Higgs or the lightest dark neutrino. 
We note that this model has rich collider signatures on dark tops, inert Higgs and $Z'$ gauge boson.
Moreover, the scalar associated to the $U(1)_D$ (and also $U(1)_{B-L}$) symmetry breaking could explain the 750 GeV diphoton excess reported by ATLAS and CMS recently.  
\end{abstract}

\maketitle


\section{Introduction}
\label{sec:intro}

The existence of neutrino masses and dark matter (DM) has been solidly confirmed by numerous observational data. 
The observation of neutrino oscillations provide evidence of non-zero neutrino masses, which cannot be explained within the standard model (SM). 
Various cosmological observations show that dark matter forms approximatively 25\% of the composition of the universe.
Therefore, both massive neutrinos and dark matter provide us very strong motivations to extend the SM.

Interestingly, it is possible that both neutrino masses and dark matter come from the same new physics at the TeV scale.
One simple and elegant realization of this idea is the scotogenic neutrino model~\cite{Ma:2006km, Kubo:2006yx, Hambye:2006zn}~\footnote{There are other radiative neutrino models~\cite{Krauss:2002px} with dark matter candidate. }. 
Similar to typical seesaw models, three right-handed neutrinos (RHNs) are introduced in this model. 
However, a $Z_2$ parity is imposed on the RHNs to forbid Yukawa interactions for neutrinos at the tree-level. 
An inert Higgs doublet $H'$~\cite{Ma:2006km, Barbieri:2006dq} is also added to provide the Yukawa coupling of the RHNs and the lepton doublets.
Neutrino masses are radiatively generated via one-loop diagrams at the TeV scale, and
the lightest neutral $Z_2$-odd particle is the DM candidate.
In this model, the dark sector includes RHNs, and an inert Higgs doublet. 
And the dark sector communicates with the SM through its Yukawa coupling to the SM lepton doublets.

In another classes of dark matter models, the so-called  flavored dark matter~\cite{Kile:2011mn, Batell:2011tc, Kamenik:2011nb, Agrawal:2011ze, Batell:2013zwa, Hamze:2014wca, Kilic:2015vka} and flavor portal dark matter~\cite{Chang:2013oia,Bai:2013iqa,Yu:2014pra}, the dark sector has the similar dark fermions and similar Yukawa interactions as the scotogenic neutrino model. 
In flavor portal dark matter models, the dark sector includes a dark particle which carries the same hypercharge as the SM fermions, and a dark matter candidate. 
The dark sector communicates with the SM via similar Yukawa coupling to the SM fermions.
Depending on the SM fermion flavor, it could be lepton flavor, quark flavor, and top flavor, etc. 
In this sense, the scotogenic neutrino model could be treated as a neutrino portal dark matter.

Therefore, it is natural to combine the two classes of models into a single framework. 
It is economic to introduce one inert Higgs doublet $H'$ for both lepton Yukawa and quark Yukawa couplings. 
In general, each flavor could have an $SU(2)_L$ singlet dark partner: $T$, $B$, $N$, $E$, which are parity-odd under a dark $Z_2$ symmetry. 
The relevant Lagrangian can be written as
\bea
	{\mathcal L}_{\rm Yuk} \simeq y_T \bar{q} \tilde{H'} T + y_B \bar{q} H' B + y_N \bar{\ell} \tilde{H'} N + y_E \bar{\ell} H' E + h.c. ,\nn\\
\eea
where $q$ and $\ell$ are SM quark and lepton doublets.
Note that the four terms in the Lagrangian corresponds to the top-flavor portal, bottom-flavor portal, neutrino-portal and lepton-portal models.
Of course, the neutrino-portal term is the Yukawa term in scotogenic neutrino model.

To further provide a unified framework, we introduce a gauged $U(1)$ symmetry in the dark sector. 
This is similar to the gauged scotogenic neutrino models~\cite{Kubo:2006rm, Adhikari:2008uc, Kanemura:2011vm, Ma:2013yga}. 
The dark fermions $T$, $B$, $N$, $E$ are charged under the gauged $U(1)$ symmetry, and SM fermions may or may not carry hidden $U(1)$ charge.  
Cancellation of the gauge anomaly determines the numbers of dark fermions and their quantum numbers.
Interestingly, after hidden $U(1)$ symmetry breaking, the dark $Z_2$ parity is still exact. 
Thus the $Z_2$ symmetry can be viewed as the remnant of the symmetry breaking, which is refered as
gauged discrete symmetry~\cite{Krauss:1988zc, Ibanez:1991pr}.

In particular, we investigate a hidden $U(1)$ model which only incorporates the up-type dark fermions: $T$, $N$, and assumes the down-type dark fermions $B$, $E$ are very heavy and thus decoupled~\footnote{
It is straightforward to extend our model to include non-decoupled down-type dark fermions. The light down-type dark fermion could enhance the 750 GeV diphoton signature.}.
Thus this model shares the common feature of the scotogenic neutrino and top flavor portal dark matter~\cite{Kilic:2015vka}.
In the original scotogenic neutrino model, if the DM candidate is the neutral component of the inert Higgs, 
most of the DM mass regions are excluded by the tension between required DM relic density 
and direct detection searches~\cite{Ma:2006km, Kubo:2006yx}. 
However, due to top favor Yukawa coupling in our hidden $U(1)$ model, this option becomes viable in this model. 
This hidden $U(1)$ can be identified as a dark $U(1)_D$ in which only dark particles carry $U(1)_D$ charge, or $U(1)_{B-L}$ gauge symmetry.
In the $U(1)_D$ case, we introduce a Dirac type dark neutrino $(N_L, N_R)$, which is different from the dark neutrinos in typical gauged scotogenic neutrino models~\cite{Kubo:2006rm, Adhikari:2008uc, Ma:2013yga}. 
Furthermore, in the gauged scotogenic neutrino models~\cite{Kubo:2006rm, Adhikari:2008uc, Ma:2013yga}, the associated $Z'$ is very hard to be produced at the LHC.
In this model, the $Z'$ could be easily produced via gluon fusion and subsequently decay exoticly. 
More interestingly, compared to the scotogenic neutrino and flavor portal DM model alone, there are richer collider and DM phenomenologies in this unified framework.
We will address the collider signatures by considering the scalar $s$ in the model as the candidate for 750 GeV diphoton resonance reported by both the ATLAS and CMS recently.

This paper is organized as follows. In Sec.~\ref{sec:model}, we discuss the mass spectrum (including radiative neutrino masses) in the hidden $U(1)_D$ model, and comment on the $U(1)_{B-L}$ model.
In Sec.~\ref{sec:darksec}, possible dark matter candidate are discussed and  its relic density and direct detection rate are calculated. 
In Sec.~\ref{sec:collider}, collider signatures of the dark sector particles are addressed. 
Then we discuss the possibility to explain the 750 GeV diphoton excess using the singlet scalar $s$ in the model in Sec.~\ref{sec:diphoton}. 
Finally we conclude this paper.


\section{The Model}
\label{sec:model}

We consider that the dark sector is gauged under a hidden $U(1)_H$ gauge symmetry. 
A discrete $Z_2$ symmetry in the dark sector is a residue symmetry of this $U(1)_H$ symmetry breaking.
Specifically, the dark sector incorporates up-type singlet fermions: heavy neutrino $N$ and heavy top $T$. 
Implementing the down-typ singlet dark fermions is very similar to the up-type $T$s.
Here we assume they are much heavier and thus decoupled. 
To connect the above dark fermions with SM particles, a second doublet $H'$ is introduced as an inert Higgs.
To be free of gauge anomaly, the dark fermions needs to assign appropriate quantum numbers. 
We present two possible charge assignments as follows: 
\bit
	\item All SM particles are not charged under $U(1)_H$, which is referred as gauged $U(1)_D$. Two heavy tops $T,T'$ and one Dirac neutrino $N$ are needed to cancel gauge anomaly. Here the Dirac heavy neutrino $(N_L, N_R)$ are different from the setup in Ref.~\cite{Kubo:2006rm, Adhikari:2008uc, Ma:2013yga}.
	\item The SM quark and leptons are charged under $U(1)_H$ with quantum numbers $\frac13$ and $-1$, which is also referred as gauged $U(1)_{B-L}$.  Two heavy tops $T,T'$ and one Majorana neutrino $N_R$ are needed to cancel gauge anomaly.  
\eit
The new particle contents and their quantum numbers are shown in Tab.~\ref{table1} and Tab.~\ref{table2}.
Two models share the same Yukawa Lagrangian, and thus essentially have very similar phenomenologies: dark matter signature, radiative neutrino mass generation, etc.  
The only differences are: (1) whether there is Dirac mass term in heavy neutrino; (2) whether the $Z'$ couples to SM fermions.
In the gauged $U(1)_{B-L}$ model, the gauge boson $Z'_{B-L}$ not only interacts with dark fermions, but also with the SM quarks and leptons. 
Therefore, to avoid stringent dilepton searches at the LHC, the gauge boson $Z'_{B-L}$ is expected to be several TeV. 
On the other hand, in hidden gauged $U(1)_D$ model, the gauge boson $Z'_D$ only interacts with dark fermions, and thus evades the strong dilepton constraints. 
In the following, we will focus on the hidden gauged $U(1)_D$ model, and only comment on the gauged $U(1)_{B-L}$ model.

{\renewcommand{\arraystretch}{1.5} 
\begin{table}[ht]
\begin{center}
\begin{tabular}{c| c c c | c c}
      &  $SU(3)_C$  & $SU(2)_L$ & $U(1)_Y$ & $U(1)_{D}$ & $Z_2$ \\ 
\hline                                                                                         
$\Phi$         & {\bf 1 }    &  {\bf 1}         &$ 0$                  & $ 2$           & $+$  \\
\hline                                                                                                    
$H'$           & {\bf 1 }    &  {\bf 2}         &$ \frac{1}{2}$        & $+1$ & $-$   \\
$N_{L,R} $          & {\bf 1 }    &  {\bf 1}          &$ 0$                 & $-1$ & $-$   \\
$ T_{L,R}$          & {\bf 3 }    &  {\bf 1}         &$+\frac{2}{3}$        & $-1$ & $-$   \\
$T'_{L,R}$          & {\bf 3 }    &  {\bf 1}         &$+\frac{2}{3}$        & $+1$ & $-$   \\
\hline
\end{tabular}
\end{center}
\caption{
The new particle contents and their quantum numbers in the $SU(2)_L \times U(1)_Y \times U(1)_{D}$ model. 
All the SM particles are not charged under $U(1)_{D}$ and even-charged under dark $Z_2$ symmetry.  
}
\label{table1}
\end{table}

{\renewcommand{\arraystretch}{1.5} 
\begin{table}[ht]
\begin{center}
\begin{tabular}{c| c c c | c c}
      &  $SU(3)_C$  & $SU(2)_L$ & $U(1)_Y$ & $U(1)_{B-L}$ & $Z_2$ \\ 
\hline                                                                                         
$\Phi$         & {\bf 1 }    &  {\bf 1}         &$ 0$                  & $ 2$           & $+$  \\
\hline                                                                                                    
$H'$           & {\bf 1 }    &  {\bf 2}         &$ \frac{1}{2}$        & $+1$ & $-$   \\
$N_{R} $         & {\bf 1 }    &  {\bf 1}          &$ 0$                 & $-1$ & $-$   \\
$ T_{L,R}$         & {\bf 3 }    &  {\bf 1}         &$+\frac{2}{3}$        & $-1$ & $-$   \\
$T'_{L,R}$          & {\bf 3 }    &  {\bf 1}         &$+\frac{2}{3}$        & $+1$ & $-$   \\
\hline
\end{tabular}
\end{center}
\caption{
The new particle contents and their quantum numbers in the $SU(2)_L \times U(1)_Y \times U(1)_{B-L}$ model. 
The SM quarks and leptons carry $\frac13$ and $-1$ under $U(1)_{B-L}$ and are even-charged under dark $Z_2$ symmetry.  
}
\label{table2}
\end{table}

In the following we investigate the hidden gauged $U(1)_D$ model contents in detail. 
The gauge symmetry is $SU(2)_L \times U(1)_Y \times U(1)_{D}$. 
This $U(1)_D$ symmetry is spontaneously broken by a singlet scalar $\Phi$:
\bea
\Phi = \frac{s + i a}{\sqrt2},
\eea 
which is only charged under $U(1)_D$. 
A residue dark $Z_2$ symmetry is left after symmetry breaking.
The dark fermions includes two vectorlike fermions $T, T'$, three generation Dirac fermions $N$. 
All the particles in the dark sector are charged under $U(1)_X$, as shown in Table~\ref{table1}.
We shall use the following parameterization of the two Higgs doublets
\bea
	H = \left(\begin{array}{c} G^+ \\ \frac{h + i G^0}{\sqrt2}\end{array}\right), \quad
	H' = \left(\begin{array}{c} h^+ \\ \frac{h' + i A}{\sqrt2}\end{array}\right).
\eea
The relevant Yukawa terms in the dark sector are 
\bea
	{\mathcal L}_{\rm Yuk} &=&  y_T \bar{q} \tilde{H}' T_R + y_N \bar{\ell} \tilde{H}' N_R  + h.c.\nn\\
	&& +  y_{T'} \bar{T'} \Phi T + y_{N'} \overline{N^c} \Phi N,
\eea
where $q$ and $\ell$ are the left-handed quark and lepton doublets in the SM, and $\tilde{H}' = i\sigma_2 H'^*$.
There are Dirac mass terms for the dark fermions 
\bea
	{\mathcal L}_{\rm mass} = m_T (\bar{T} T  + \bar{T'} T') + M_D \overline{N}  N
\eea
In the scalar sector, we could write down the most general scalar potential  
\bea
	V_{\rm scalar} &=& -\mu_s^2 |\Phi|^2 + \lambda_s |\Phi|^4  + V_{H,H'}  \nn \\
	&+&  \lambda_{h} |H|^2|\Phi|^2  + \lambda_{h'} |H'|^2 |\Phi|^2,
\eea
where $V_{H,H'} $ is the Higgs potential for the inert Higgs doublet and the SM Higgs 
\bea
	V_{H,H'}  &=& \mu_1^2 |H|^2 + \mu_2^2 |H'|^2 + \lambda_1 |H|^4 + \lambda_2 |H'|^4   \\
	&+&   \lambda_3 |H|^2 |H'|^2  +  \lambda_4 |H^\dagger H'|^2  
	+ \frac{\lambda_5}{2} \left[ (H^\dagger H')^2 + h.c.\right].\nn
\eea

According to the above scalar potential, both the SM Higgs $H$ and $\Phi$ obtain vacuum expectation values (vev):
\bea
	\langle H \rangle = v, \quad \langle \Phi \rangle = u. 
\eea
There is no vev for the scalar doublet $H'$ and thus the dark $Z_2$ symmetry is exact.
After spontaneous symmetry breaking, the $U(1)_D$ symmetry is broken down to a residue $Z_2$ symmetry: $Z_2(\Phi) = +1$, and $Z_2(T, T', N, H') = \pm 1$. 
We identify the dark $Z_2$ symmetry as a remnant of the $U(1)_D$ symmetry breaking with $Z_2(T, T', N, H') = - 1$.
Therefore, after symmetry breaking the dark $Z_2$ symmetry is still exact and thus stabilize the dark matter candidate.

The symmetry breaking induces the scalar masses in the dark sector:
\bea
	m_{h'}^2  &=& \mu_2^2 + \frac{\lambda_3 + \lambda_4 + \lambda_5}{2} v^2,\nn\\
	m_{A}^2   &=& \mu_2^2 + \frac{\lambda_3 + \lambda_4 - \lambda_5}{2} v^2, \nn\\
	m_{h^\pm} &=& \mu_2^2 + \lambda_3 v^2.
\eea
Because there is no mixing between $Z$ and $Z'$, the $Z'$ mass is
\bea
	m_{Z'} = \frac12 g_D u,
\eea
where $g_D$ is the gauge coupling strength of the $U(1)_D$.
Furthermore, new particles with the same charge mix together after symmetry breaking.
For the real components of the scalars $(h, s)$, we obtain
\bea
	{\mathcal M}^2_{\rm S} = \left(\begin{array}{cc} 2 \lambda_1 v^2 & \lambda_h v u\\ 
	\lambda_h v u & 2 \lambda_s u^2 \end{array}\right),
	\label{eq:scalarmass}
\eea 
The mass eigenstates and mixing matrix are
\bea
	\left(\begin{array}{c} h_1 \\ h_2 \end{array}\right) &=& R (\phi) \left(\begin{array}{c} h  \\ s \end{array}\right), \quad R (\phi) =  \left(\begin{array}{cc} \cos\phi  &  -\sin\phi\\ 
	\sin\phi & \cos\phi \end{array}\right).
\eea
where the mixing angle and mass eigenvalues are
\bea
\tan2\phi &=& \frac{\lambda_h v u}{ \lambda_1 v^2 - \lambda_s u^2 } ,\\
m^2_{1,2} &=& 2\lambda_1 v^2 + 2\lambda_s u^2 \mp \sqrt{ (\lambda_1 v^2 - \lambda_s u^2 )^2 + \lambda_h^2 v^2 u^2 }.\nn
\eea
Similarly, for the heavy vectorlike fermions $(T, T')$, we read
\bea
	{\mathcal M}_{\rm F} = \left(\begin{array}{cc} m_T  & M_T\\ 
	M_T &  m_T \end{array}\right),
\eea 
where $M_T = \frac{y_{T'} u}{\sqrt2}$.
The mass eigenstates and eigenvalues are
\bea
	\left(\begin{array}{c} T_{1} \\ T_{2} \end{array}\right)_{L,R} = R({\theta_{L,R}})\left(\begin{array}{c} T  \\ T' \end{array}\right)_{L,R},
	m_{T_{1,2}} = M_T \mp m_T.\nn\\
\eea
Note that two mixing angles are not independent, with a relation
\bea
	\tan\theta_R = \frac{m_{T_1}}{m_{T_2}} \tan\theta_L.
\eea
If the Dirac mass term $m_T \ll M_T$, the mas eigenstates $T_{1,2}$ is almost degenerate with $\theta_L \simeq \theta_R$.

After symmetry breaking, the dark neutrino $N$ becomes pseudo-Dirac fermion due to Majorana mass terms.
The mass matrix $(N_L, N_R)$ is written as
\bea
	{\mathcal M}_{\rm N} = \left(\begin{array}{cc}  m_R  & M_D\\ 
	M_D & m_R \end{array}\right),
	\label{eq:neutrinomass}
\eea 
where $m_R = \frac{y_{\Phi} u}{\sqrt2}$, and the mass eigenstates are 
\bea
\left(\begin{array}{c} {\mathcal N}_{+} \\ {\mathcal N}_{-} \end{array}\right) &=& R (\alpha) \left(\begin{array}{c} N_L  \\ N_R \end{array}\right), \quad m_{N_{\mp}} =  M_D \mp m_R.
\eea
Majorana neutrino masses in this model are generated at one-loop via the exchanges of $h'$ and $A$, which is the radiative seesaw mechanism~\cite{Ma:2006km}. 
In this mechanism, the one-loop induced dimension-six operator $\ell \ell \Phi H H$, 
and the resulting neutrino mass matrix is given by~\cite{Ma:2006km}
\bea	
m_{\nu_{ij}} = \sum_k \frac{y_{ik} y_{jk} m_{R_k}}{16 \pi^2} \left[{\mathcal I}\left(\frac{m_{h'}^2}{m_{R_k}^2}\right) - {\mathcal I}\left(\frac{m_{A}^2}{m_{R_k}^2}\right)\right],
\label{eq:nvmass}
\eea
where $i,j,k$ are the generation index and the loop function is defined as
\bea
	{\mathcal I}(x) = \frac{x}{x-1} \ln x.
\eea
Here we denote
\bea
y_{ik} \equiv (y_N)_{ik}, \quad m_{R_k} \simeq \frac{(y_{\Phi})_k u}{\sqrt2}.
\eea
To obtain a small neutrino mass, the mass splitting between $h'$ and $A$ should be small, and thus
a small $\lambda_5$ is needed.
If we further demand $m_{R_k} \ll m_A$, the Eq.~\ref{eq:nvmass} is approximated by a seesaw formula
\bea
	m_{\nu_{ij}} \sim \sum_k  y_{ik} y_{jk}  \frac{\lambda_5 v^2}{8 \pi^2}  \frac{m_{R_k}}{m_A^2}.
\eea 
For a TeV scale vev $u$, given $\lambda_5 \sim y_{\Phi} \sim 10^{-2}$, the 0.1 eV neutrino mass could be realized. Since the flavor structure is quite similar to the standard seesaw mechanism, the predicted neutrino oscillation matches with the observed data~\cite{Ma:2006km, Kubo:2006yx, Hambye:2006zn}.

In summary, the particle contents in the mass eigenstates are 
\bit
\item new gauge boson $Z'$ and a heavy real scalar $s$ associated to the $U(1)_X$ symmetry breaking;
\item in dark sector, two colored dark fermions $T_{1,2}$ (dark tops), two pseudo-Dirac dark neutrinos ${\mathcal N}_{\pm}$ for each generation, inert doublet components $h'$, $A$, and $h^\pm$.   
\eit
The dark matter candidate could be $h'$ ($A$) or the lightest ${\mathcal N} \equiv {\mathcal N}_-$.


\section{Dark Matte Candidate}
\label{sec:darksec}

From Sec~\ref{sec:model}, we note that the mass hierarchy: $m_{h'} \simeq m_A \gg m_R$ is favored to obtain the radiative neutrino mass. 
Furthermore, to determine the dark matter candidate, the Dirac mass $M_D$ in Eq.~\ref{eq:neutrinomass} plays an important role. 
In the pattern $m_{h'} \simeq m_A < M_D$, the DM candidate will be $h'$ or $A$ depending on the sign of the $\lambda_5$. 
On the other hand,  if $m_{h'} \simeq m_A > M_D$, DM will be the lightest ${\mathcal N}$. 
When the Dirac mass $M_D \simeq m_R$, which corresponds to very light ${\mathcal N_-}$, the lightest ${\mathcal N}$ could be a KeV warm dark matter.

If the neutral component $h'$ ($A$) of the inert Higgs  is the DM, its phenomenology is very similar to the inert doublet Higgs (IDHM) model~\cite{Ma:2006km,Barbieri:2006dq,Cao:2007rm}. 
It has been known that the DM in the IDHM is very constrained~\cite{Goudelis:2013uca}.
The LHC measurements on the Higgs invisible width exclude the parameter region with $m_{\rm DM} < m_{h}/2$, while the direct detection experiments exclude the immediate DM mass $100 \,\,{\rm GeV} \lesssim m_{\rm DM} \lesssim 850$ GeV. 
The DM mass regions which could evade current direct detection and LHC constraints are 
\bit
\item Higgs resonance region $m_{\rm DM} \simeq m_{h}/2$, in which DM particles annihilate into a resonant s-channel Higgs boson.
\item off-shell $W^+ W^-$ region $m_{\rm DM} \lesssim 100$ GeV, in which DM particles annihilate into off-shell $W^+ W^-$ pairs. This region will be constrained by next generation direct detection experiments. 
\item scalar resonance region $m_{\rm DM} \simeq m_{s}/2$, in which DM particles annihilate into a resonant s-channel heavy $s$ boson.
\item high mass region $m_{\rm DM} \gtrsim 850$ GeV.
\item immediate mass region $m_{\rm DM} \gtrsim 175$ GeV could be allowed in very narrow parameter space if the DM couplings to the Higgs boson is small~\footnote{The relic density could be realized through top flavored dark matter via t-channel $h' h' (A A) \to t\bar{t} $, while the direct detection constraints are loose~\cite{Kilic:2015vka}. }, or there are deconstructive cancellation in direct detection~\footnote{In DM-nucleon scattering, the contribution from the loop-induced photon exchange could have opposite sign of the contribution from the the Higgs exchange diagrams~\cite{Hamze:2014wca}. Therefore, the direct detection constrains could be evaded. }.  
\eit
It is very interesting to see that if the coupling of the Higgs boson to the DM is small, 
the top flavor portal coupling will dominate the relic density, but immediate mass $h'$ ($A$) is still allowed by current LUX searches. 
In this case, the DM phenomenology is exactly the same as the top-flavored dark matter model~\cite{Kilic:2015vka}. So we won't discuss it here and refer to Ref.~\cite{Kilic:2015vka} for details.

Another possibility is that  the DM is the lightest ${\mathcal N}$. Depending on the values of the Dirac mass $M_D$, the DM mass could be in a broad range. 
When the DM is very light (such as KeV scale), it becomes warm dark matter.
In this work, we won't consider this special case, and will only focus on cold dark matter scenario. 
The cold DM annihilation processes are 
\bea
	t{\rm -channel}: &\,\,& {\mathcal N} {\mathcal N} \to \ell^+ \ell^-, {\mathcal N} {\mathcal N} \to \nu \bar{\nu}.\nn\\
	s{\rm -channel}:  &\,\,& {\mathcal N} {\mathcal N} \to s(h) \to {\rm SM} \,{\rm \overline{SM}}.
\eea
The $s$-channel process depends on the couplings $y_{N'}$, which are taken to be very small to explain the neutrino mass. 
So the dominant channel should be the $t$-channel process. 
The thermal averaged $t$-channel cross section is
\bea
	\langle \sigma v_{\rm rel} \rangle_{\rm t-chan} \simeq \frac{\sum_{ij}|y_{i1}^2 y^2_{j1}|^2}{48 \pi M_{{\mathcal N}}^2}  \left(\frac{ 1+ x^2}{(1 + x)^4} + \frac{ 1+ y^2}{(1 + y)^4} \right) v_{\rm rel}^2,\nn\\
\eea
where $x = m_A^2/M_{{\mathcal N}}^2$ and $y = m_{h^+}^2/M_{{\mathcal N}}^2$.
Note that the thermal cross section is  $p$-suppressed, which implies not so small coupling $y_{N}$ is needed to obtain required relic density.
On the other hand, the coupling $y_{N}$ is restricted by constraints from lepton flavor violation processes, such as $\mu \to e \gamma$ and $\tau \to \mu \gamma$. 
In the limit $m_{R_k} \ll m_A$, the branching ratio is
\bea
{\rm Br}(\ell_\alpha \to \ell_\beta \gamma)\sim \frac{\alpha_e}{64\pi}\left(\sum_k y_{\alpha k}y_{\beta k}\right)^2 \frac{v^4}{m_A^4},
\eea
In general, there is a strong tension between the relic density requirement and the upper limit from lepton flavor violation processes~\cite{Kubo:2006yx}. 
However, there are parameter regions which could evade these constraints:
\bit
\item special flavor textures~\cite{Kubo:2006yx, Ibarra:2016dlb}. One special case is the degenerate Majorana masses $m_{R}$ for dark neutrinos~\cite{Kubo:2006yx}. In this case, co-annihilation in the DM annihilation exists.
\item $s(h)$ resonance region $m_{{\mathcal N}} \simeq m_{s(h)}/2$, in which DM particles annihilate into a resonant s-channel $s(h)$ boson~\cite{Suematsu:2010gv}.  
\eit

\begin{figure}
  \includegraphics[width=0.36\textwidth]{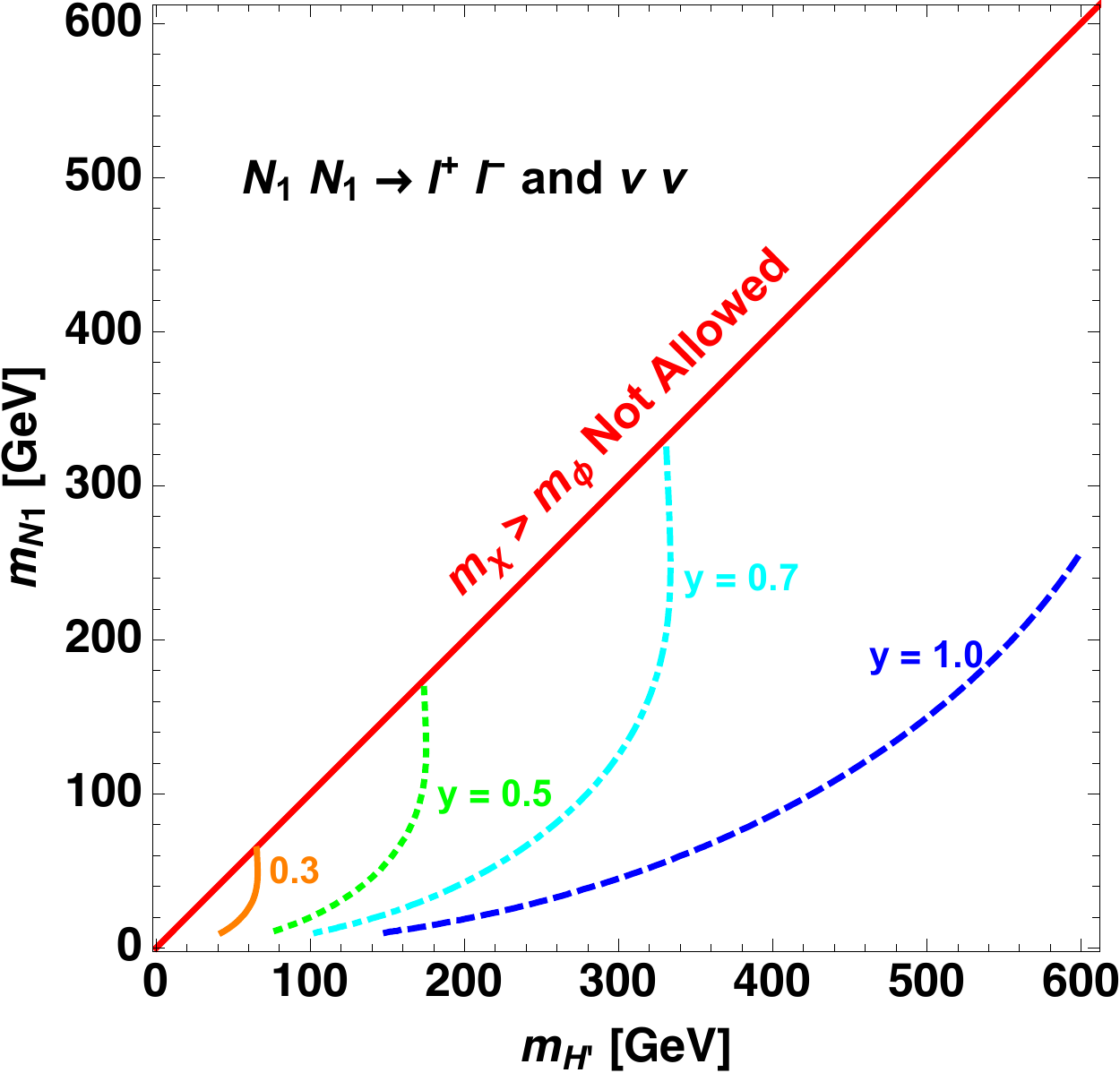}
\caption{\label{fig:relic} The contours for several values of the coupling $y_N$ which gives rise to the required relic density in the $(m_{H'}, m_{{\mathcal N}})$ plane. 
Here we take $m_{H'} = m_{h'} = m_{A}$.}
\end{figure}

In this work we focus on the case ${\mathcal N}$ as the dark matter candidate. 
In Fig.~\ref{fig:relic}, we show the relic density contours in the $(m_A, m_{{\mathcal N}})$ plane for different $y \equiv y_{ik}$.  
As discussed in Ref.~\cite{Kubo:2006yx,Suematsu:2010gv}, given degenerate dark neutrino masses, the required relic density could be satisfied while keeping the lepton flavor violation rate under the experimental limits.
We also examine the direct detection constraints on the parameter space. 
Similar to the relic density, the interesting region is small coupling $y_{N'}$. 
This implies that the DM-nucleon scattering at the tree-level via the $s(h)$ portal is highly suppressed. 
We expect the loop-induced DM-nucleon process dominates the direct detection, as shown in Fig.~\ref{fig:dd2}. 
We know that the lightest ${\mathcal N}$ is the mixture of Dirac fermion and Majorana types. 
The Majorana component will only contribute to spin-dependent cross section, while the 
Dirac component contributes to spin-independent cross section. 
From requirement of obtaining radiative neutrino mass, the Majorana component of the lightest ${\mathcal N}$ 
is much smaller than the Dirac component.
Since the direct detection constraint on the spin-dependent cross section is very loose~\cite{Ibarra:2016dlb}, we will only care about direct detection constraint on the Dirac component of the ${\mathcal N}$.
The one-loop induced Dirac DM-nucleon interaction term is
\bea
  {\mathcal L}_{\rm eff} &=& b_{{\mathcal N}}  \overline{{\mathcal N}} \gamma_\nu   {\mathcal N} \partial_\mu F^{\mu\nu} +
  \mu_{{\mathcal N}}  \overline{{\mathcal N}} i\sigma_{\mu\nu}   {\mathcal N} F^{\mu\nu},
\label{eq:gammafdm2}
\eea
where $\mu_{{\mathcal N}}$ and $b_{{\mathcal N}}$ are given by
\bea
    \mu_{\mathcal N} &=& -\frac{ie \lambda^2}{64\pi^2 } \int^1_0 d y\  2 m_{\mathcal N}  \frac{ y(1-y)}{(1-y)m_A^2 -  y(1 - y) m_{\mathcal N}^2} ,\\
    b_{\mathcal N} & = & 		-\frac{ie \lambda^2}{64\pi^2 } \int^1_0 d y\  \frac{1}{6}\frac{ 3y^3 - 9y^2 +2}{(1-y)m_A^2 -  y(1 - y) m_{\mathcal N}^2}.
\eea
In the limit $m_R \ll M_D$, the lightest ${\mathcal N}$ is almost Dirac type, and thus should obtain the strongest bounds from direct detection. 
Following Ref.~\cite{Hamze:2014wca}, we calculate the spin-independent cross section using effective Lagrangian in Eq.~\ref{eq:gammafdm2}. 
Fig.~\ref{fig:lux} shows that given the coupling size, the strongest direct detection constraints from the LUX experiment~\cite{Akerib:2013tjd}. 
From Fig.~\ref{fig:lux}, we note that the direct detection constraints are not so tight, and there are still large allowed parameter regions.

\begin{figure}
  \includegraphics[width=0.2\textwidth]{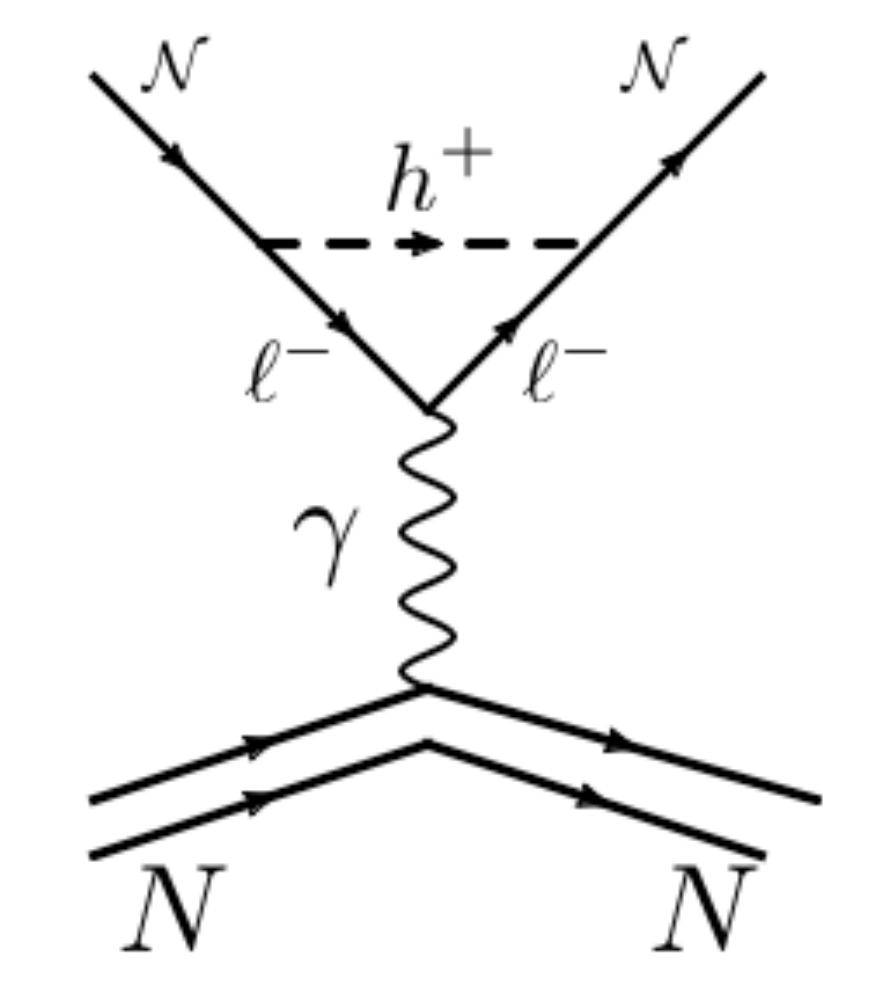}
  \includegraphics[width=0.2\textwidth]{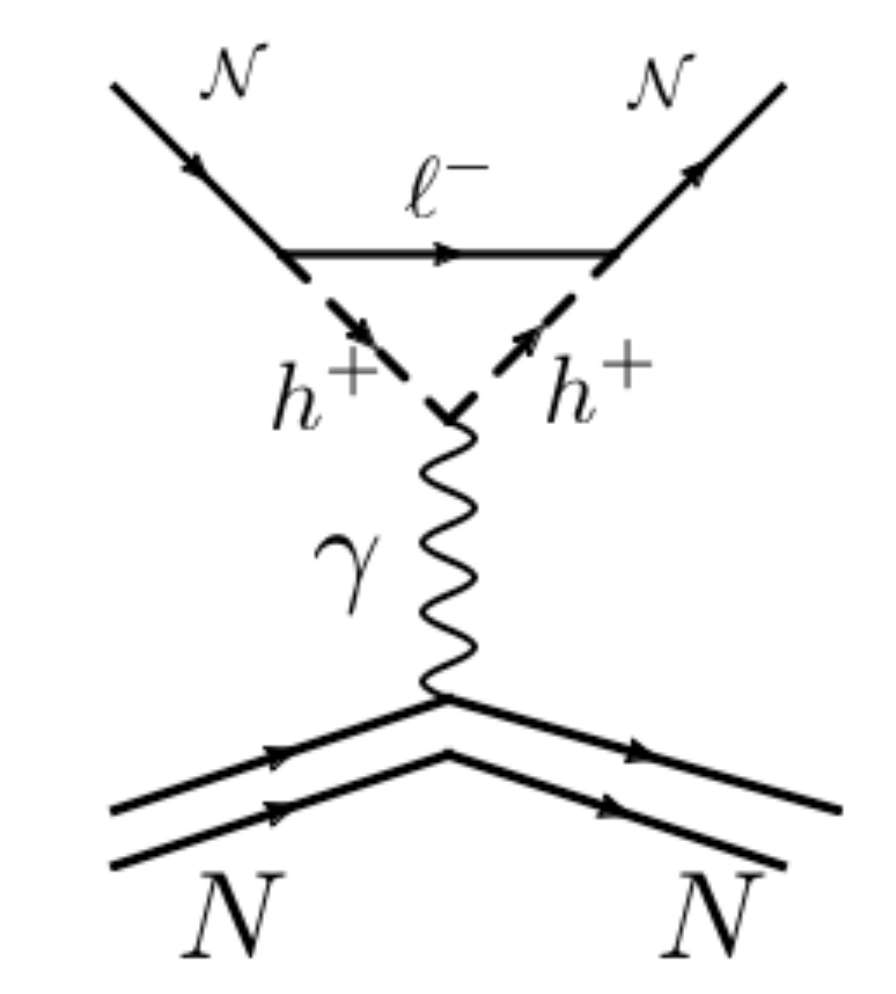}
\caption{\label{fig:dd2}  Loop-induced Feynman diagrams on dark matter-nucleon (N) scattering via photon exchange.}
\end{figure}

\begin{figure}
  \includegraphics[width=0.36\textwidth]{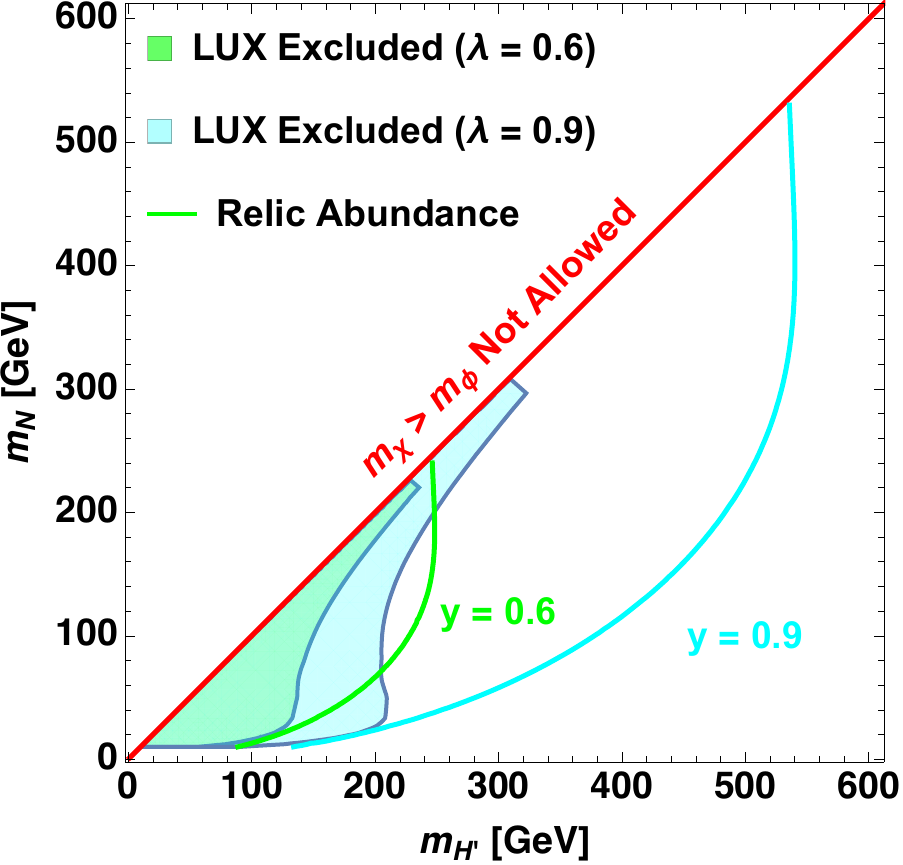}
\caption{\label{fig:lux}   The blue and cyan regions on $(m_{H'}, m_{{\mathcal N}})$ are the excluded regions  with $y_N = 0.6, 0.9$ by the LUX experiment. }
\end{figure}


\section{Collider Searches on Dark Sector}
\label{sec:collider}

The dark sector encounters not only the DM constraints, but also experimental constraints from high energy colliders, such as LEP and LHC. 
The electroweak precision data from LEP put constraints on dark particles in $SU(2)_L$ non-singlet, such as inert doublet components $A$, $h'$ and $h^\pm$. 
In this model, the LEP constraint is quite similar to the IDHM~\cite{Barbieri:2006dq}. 
According to Ref.~\cite{Barbieri:2006dq}, the S parameter contribution is negligible  and 
the $T$ parameter approximates
\bea
	\Delta T \simeq \frac{1}{24 \pi^2 \alpha v^2} (m_{h^\pm} - m_A)(m_{h^\pm} - m_{h'}).
\eea
The recent experimental limits on the $S, T$ parameters (given $U=0$) are~\cite{Baak:2011ze}
\bea
	S = 0.06 \pm 0.09, \quad T = 0.10 \pm 0.08.
\eea
Therefore, the small mass splitting between $h^\pm$ and $h'$ ($A$) is favored by the LEP data.

\begin{figure}
  \includegraphics[width=0.25\textwidth]{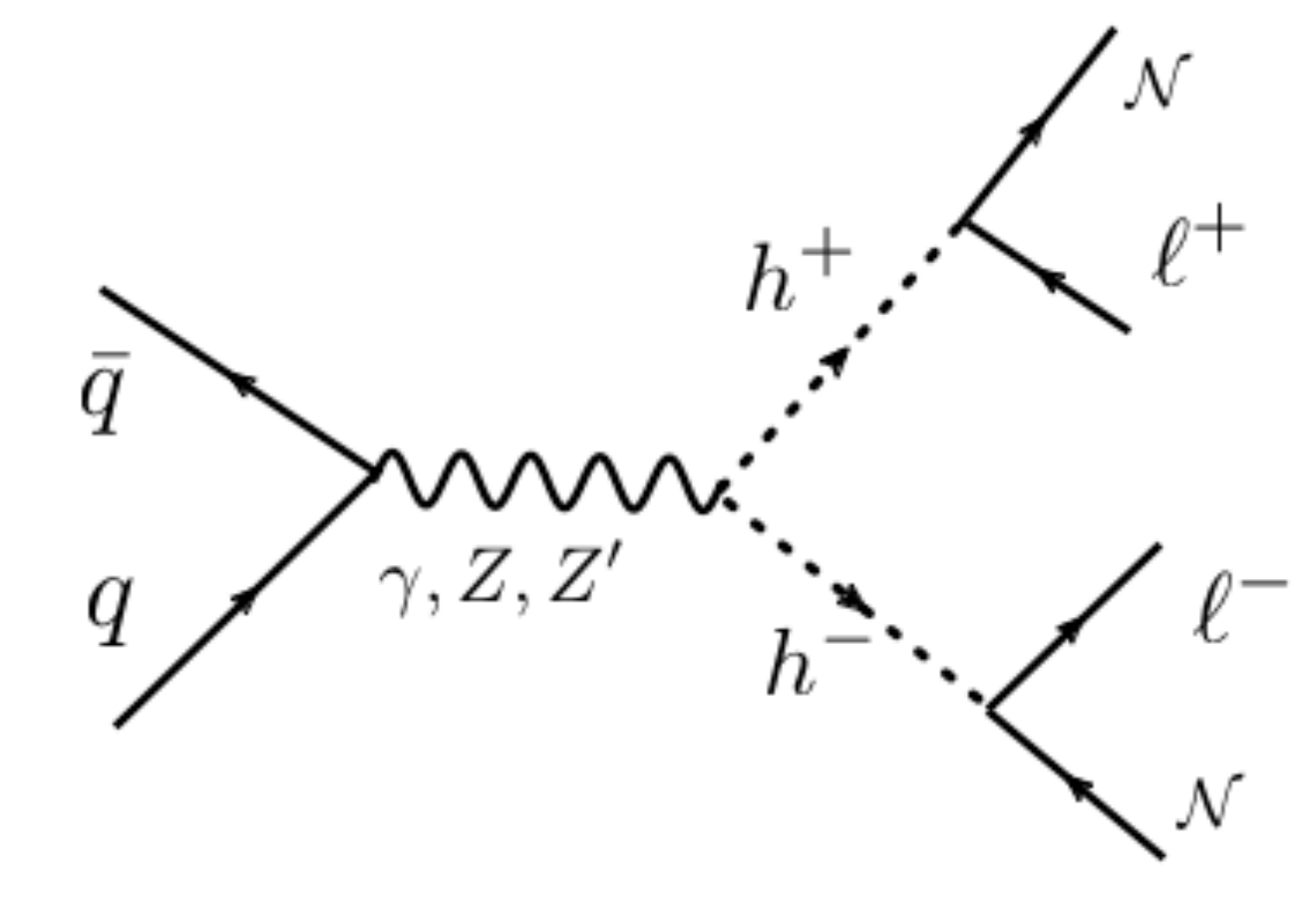}
  \includegraphics[width=0.3\textwidth]{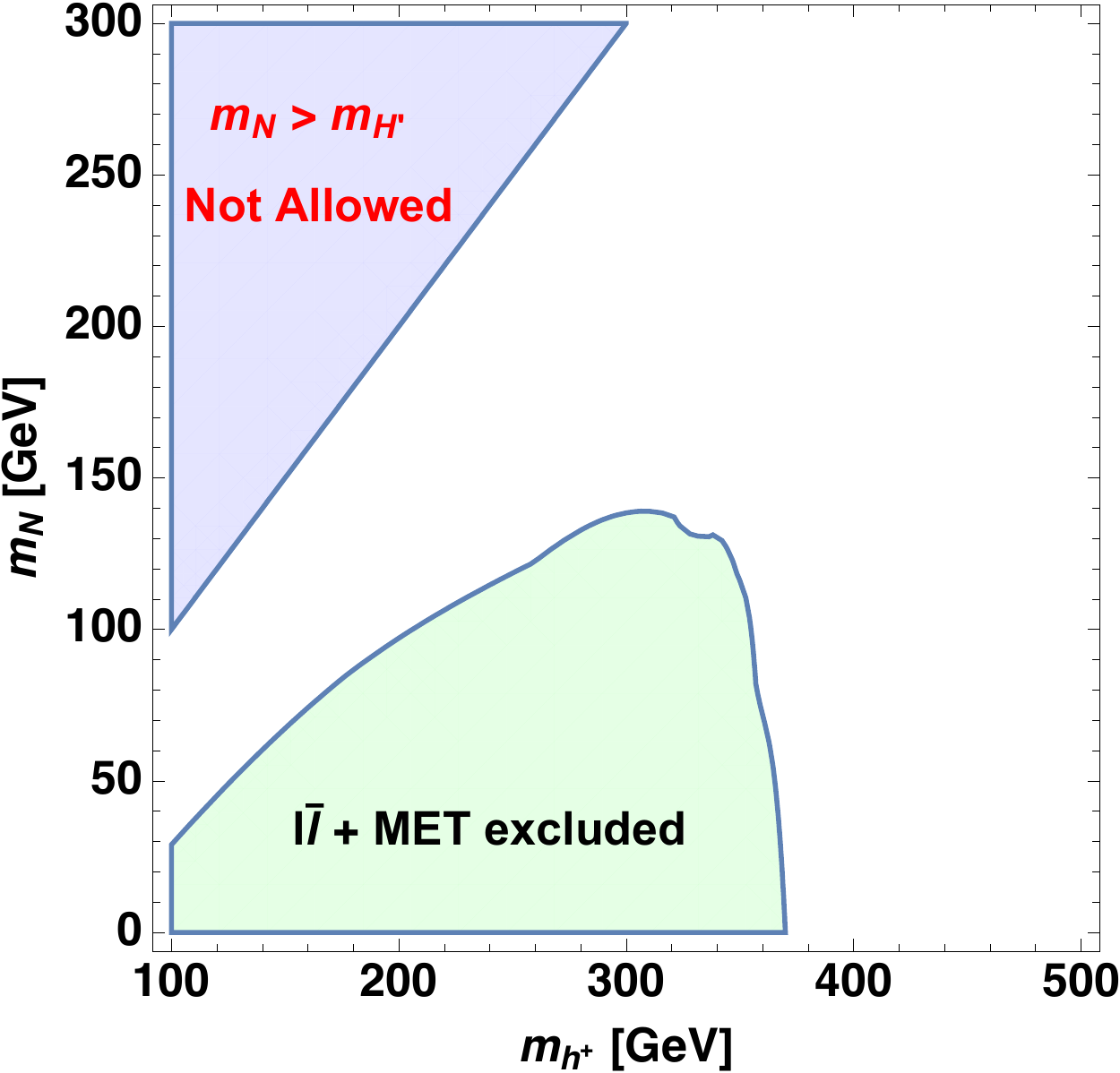}
\caption{\label{fig:chargeH}   (Upper panel) The Feynman diagram for the $h^\pm$ pair production and dominant decay channel. (Lower panel) The excluded parameter region on $(m_{h^\pm}, m_{\mathcal N})$ by the LHC dilepton plus missing energy searches.}
\end{figure}

Since we focus on the case that the lightest ${\mathcal N}$ is the DM candidate, 
we expect that the collider signatures of
inert doublet components are different from signatures in IDHM~\cite{Cao:2007rm, Goudelis:2013uca}. 
Assuming the compressed masses among $A$, $h'$ and $h^\pm$,
the dominant production and decay channels of the inert Higgs components are
\bea
	pp \to h' h' (AA) \to \nu {\mathcal N} +  \nu {\mathcal N},  	
\eea
which gives rise to the invisible final states, and 
\bea
	pp \to h^+ h^- \to \ell^+ {\mathcal N} +  \ell^- {\mathcal N},	
\eea
which exhibits the dilepton plus transverse missing energy (MET) final states.
At the LHC, there are searches on these final states. 
And the tightest constraint comes from dilepton plus MET channel~\cite{Aad:2014vma}. The Feynman diagram is shown in Fig.~\ref{fig:chargeH} (upper panel).
The experimental limit can be recasted from the exclusion limits~\cite{Aad:2014vma} on SUSY chargino/slepton searches at the 8 TeV LHC. 
In the ATLAS search~\cite{Aad:2014vma}, two scenarios are considered: the chargino pair is produced and subsequently decays to neutralino and a $W$ boson, and the slepton are pairly produced and then each decays to dilepton plus neutralino. 
We recast the slepton $\tilde{\ell}_L$ exclusion limit and obtain the exclusion contour in the $(m_{h^\pm}, m_{\mathcal N})$ plane, shown in Fig.~\ref{fig:chargeH} (lower panel).
From the Fig.~\ref{fig:chargeH} we see the upper limits on $m_{h^\pm}$ is around 370 GeV depending on the DM mass. 
The constraint on $(m_{h^\pm}, m_{\mathcal N})$ is  loose and there are still large allowed parameter space.


\begin{figure}
  \includegraphics[width=0.25\textwidth]{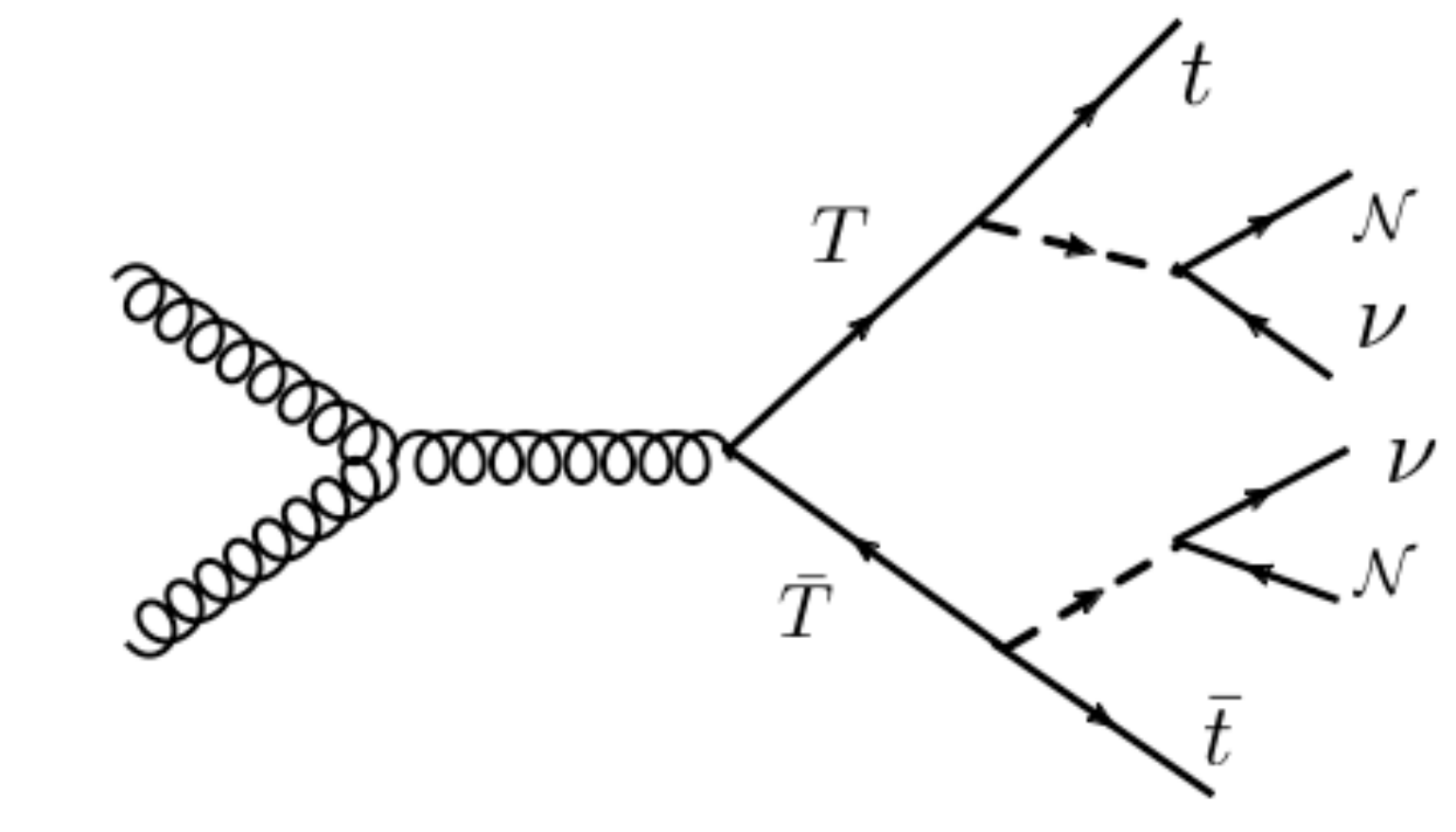}
  \includegraphics[width=0.3\textwidth]{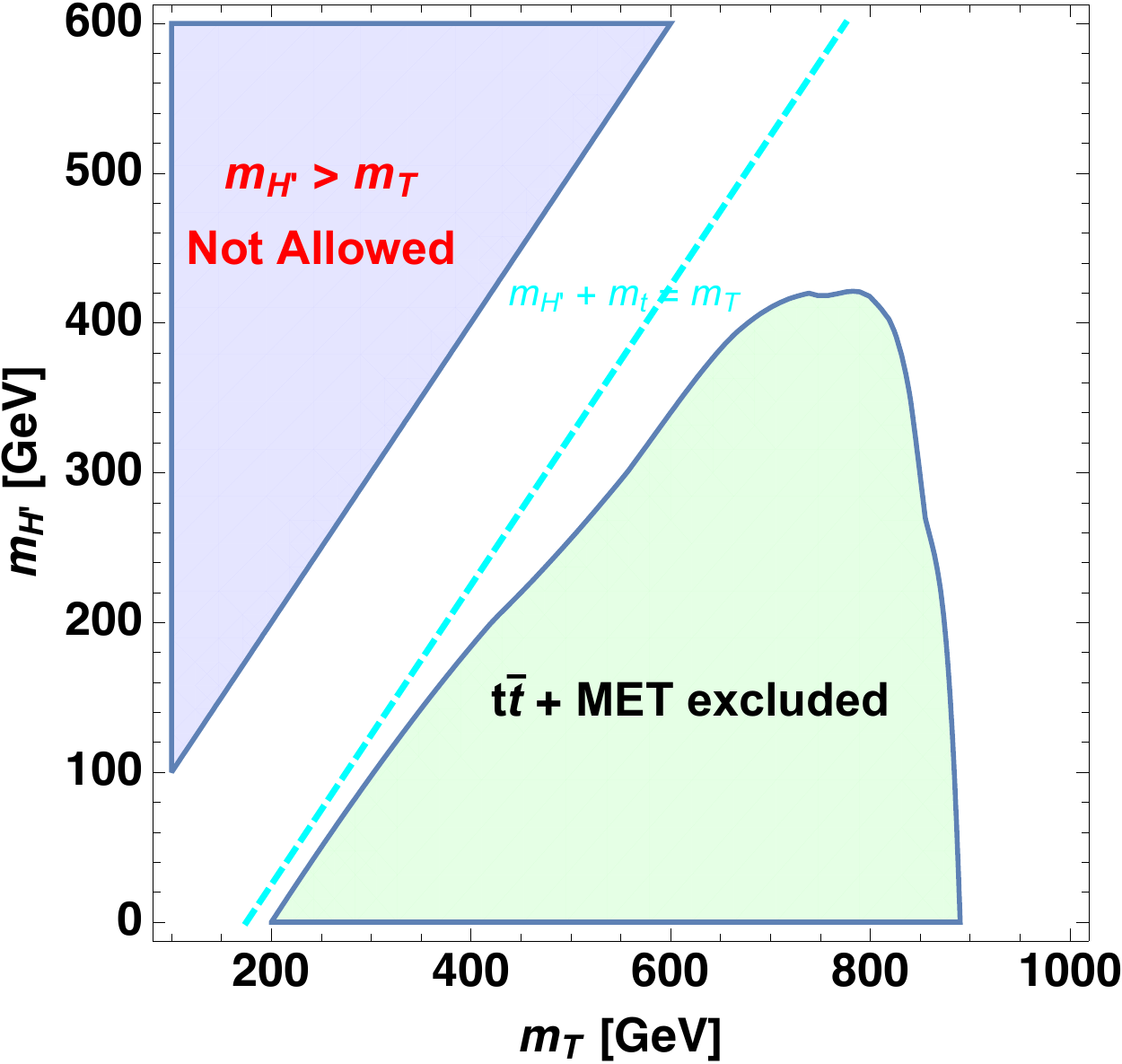}
\caption{\label{fig:heavyT}    (Upper panel) The Feynman diagram for the $T$ pair production and dominant decay channel. (Lower panel) The excluded parameter region on $(m_{T}, m_{h'})$ by the LHC hadronically decaying top pair plus missing energy searches.}
\end{figure}

Unlike the charged Higgs, the dark vectorlike tops $T$ and $T'$ encounter stronger constraints at the current LHC searches due to its QCD production cross section at the LHC.  
The dominant production and decay channels of the dark tops are
\bea
	p p &\to& T \bar{T} \to t h' + \bar{t} h' \to t {\mathcal N} \nu + \bar{t} {\mathcal N} \nu, \\
	p p &\to& T \bar{T} \to b h^+ + \bar{b} h^- \to b\ell^+ {\mathcal N} +  \bar{b}\ell^- {\mathcal N}.
\eea 
In these two processes, the final states are the top pair plus MET, and bottom/lepton pair plus MET separately. 
The Feynman diagram for the top pair plus MET is shown in Fig.~\ref{fig:heavyT} (upper panel). 
The same final states have been searched at the LHC: the stop pair production and decays~\cite{Aad:2015pfx}. 
The combined exclusion limit with 20 fb$^{-1}$ data at the 8 TeV LHC are summarized in Ref.~\cite{Aad:2015pfx}. 
From the exclusion limits, we see that the all hadronic top final states~\cite{Aad:2014bva} put the tightest constraints on the stop quark masses. 
We recast the experimental limit~\cite{Aad:2014bva} through a rescale of the total rates and cut efficiencies.
Note that $m_{h'} > m_{\mathcal N}$, we obtain the exclusion limits on $(m_T, m_{h'})$, instead of $(m_T, m_{\mathcal N})$.
Fig.~\ref{fig:heavyT} (lower panel) shows the excluded region on $(m_T, m_{h'})$ plane. 
We see that the dark tops could be excluded up to around 890 GeV. However, there is still a small region that the dark tops could be light: $0 <m_T - m_{h'} < m_t$. 
This is referred as the stealth region. 
We will discuss collider signatures in this region in sec.~\ref{sec:diphoton}.

The unique signature of this gauged hidden $U(1)$ model is the $Z'$ signature. 
If the gauge symmetry is $U(1)_{B-L}$, since both quarks and leptons carry $B-L$ charge, 
the $Z'$ mainly decays to SM quarks and leptons. 
From the exclusion limits on the $Z'$ searches at the LHC~\cite{Aad:2014cka}, 
the $Z'$ should be heavier than 2.5 TeV.
If the gauge symmetry is $U(1)_D$, since the SM fermions carry no dark charge, 
the production and decay of the $Z'$ is quite different from the typical $Z'$. 
Although the inert Higgs could induce small mixing between $Z$ and $Z'$, 
the $Z'$ couplings to the SM fermions via the $Z-Z'$ mixing is suppressed.
Thus the feasible production channel of the $Z'$ is through gluon fusion process with dark tops in the loop.
The dominant decay channels could be 
\bea
	Z' \to T \bar{T}, h^+ h^-, h' h', A A, {\mathcal N}{\mathcal N}
\eea
depending on the mass hierarchy between $m_{Z'}/2$ and other dark particles.
These exotic channels for the $Z'$ searches give rise to jets/leptons plus MET final states.
Compared to the dilepton final states, the exclusion limit on this hidden $Z'$ should be much weaker than the dileptonic $Z'$.  
From the jets/leptons plus MET searches, it is still fine to have the $Z'$ mass around several hundred GeVs.

The production and decays of the scalar $s$ is very interesting. 
Depending on the mixing angle $\sin\phi$, $s$ could have quite different decay patterns.
We note that the mixing angle $\sin\phi$ are constrained by the Higgs coupling measurements at the LHC. 
The current limit on $\sin\phi$ is around $\sin\phi \lesssim 0.4$~\cite{Xiao:2014kba}. 
Thus one expect that $s$ dominantly decays to SM particles through mixing with the SM Higgs boson.
Depending on the mass hierarchy between $m_s/2$ and dark fermions, there are other decay channels  
\bea
s \to T \bar{T}, h^+ h^-, h' h', A A, {\mathcal N}{\mathcal N},
\eea
similar to the gauge boson $Z'$ decays. 
If all the dark particles except ${\mathcal N}$ are heavier than $m_s/2$, and the mixing to the SM Higgs boson is small, the dominant decay channels of the $s$ are
\bea
	s \to gg, \gamma\gamma, {\mathcal N}{\mathcal N},
\eea
where $gg, \gamma\gamma$  channels are loop-induced decay channels and the ${\mathcal N}{\mathcal N}$ channel is the invisible decay at tree-level. 
One may expect that the tree-level invisible decay dominates over the loop-level decay. 
In fact, to give rise to radiative neutrino mass, the coupling of the $s$ to DM needs to be very small.
This suppresses the tree-level decay and enhances the loop-induced decay channels, such as $gg, \gamma\gamma$ channels.
We will discuss the possible signatures in next section.


\section{Diphoton Signature at the LHC}
\label{sec:diphoton}

As discussed above, the scalar $s$ could dominantly decay to digluon and diphoton if dark tops  and components of the inert Higgs are heavier than $m_s/2$ and the mixing between $s$ and the SM Higgs is very small.  
%
%
In this particular parameter region, it is very interesting that the LHC could observe diphoton signature.

Recently  both the ATLAS and CMS experiments at the LHC observed a resonance near 750 GeV in its diphoton final states~\cite{ATLAS13, CMS13}. On the other hand, this diphoton excess is not accompanied by other channels, such as $WW$, $ZZ$, $\ell\ell$, $jj$ final states. Of course, this could just be statistical fluctuation due to limited integrated luminosity at the 13 TeV LHC. If it is taken to be new physics signature~\cite{Franceschini:2015kwy}, the excess corresponds to production and decays of a spin-0 or spin-2 resonance with $5\sim 10$ fb rate.

\begin{figure}
  \includegraphics[width=0.3\textwidth]{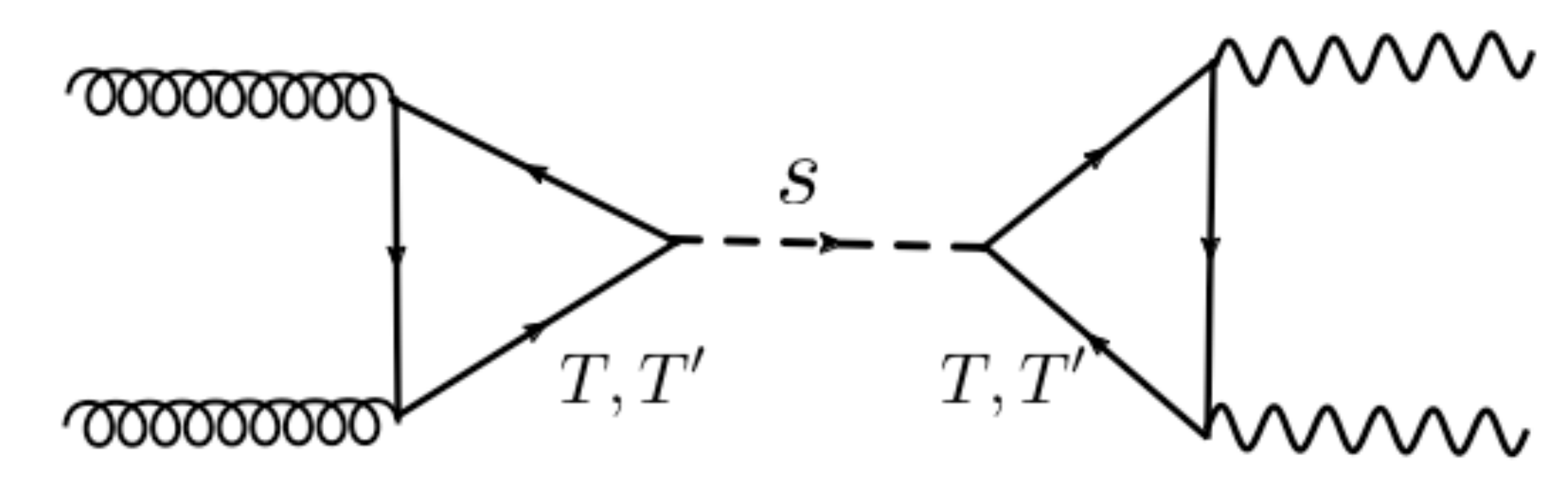}
  \includegraphics[width=0.3\textwidth]{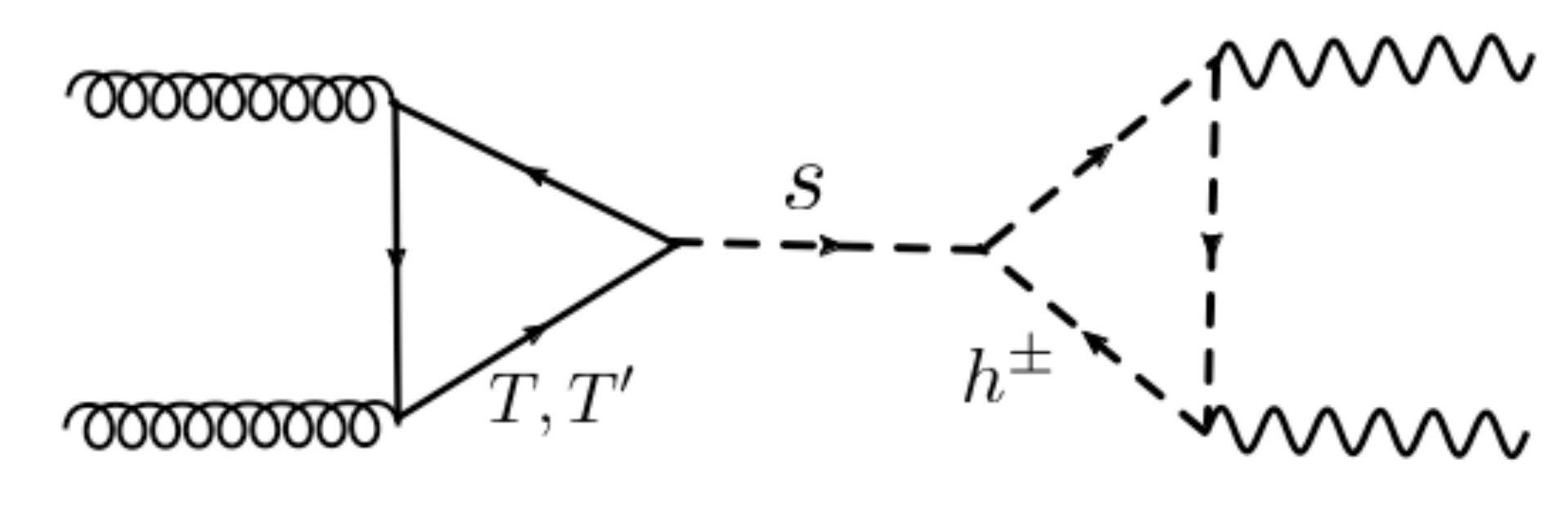}
\caption{\label{fig:diphoton} The  Feynman diagrams for the production and diphoton decay of the scalar $s$. 
For the gluon fusion production, the dark tops $T (T')$ are running in the loop. For the diphoton decay, both the dark tops (upper diagram) and the charged inert Higgs (lower diagram) are running in the loop.}
\end{figure}

We would identify the scalar $s$ as the 750 GeV resonance. This scalar $s$ is responsible to the symmetry breaking of the hidden gauged $U(1)$. Similar ideas on identifying 750 GeV scalar as the gauged $U(1)$ remnant are also considered in Ref.~\cite{Das:2015enc}. When the $s-h$ mixing is small, $s$ is expected to be produced mainly via gluon fusion with colored dark fermions running in the loop. It will subsequently decays to either dark fermions, or digluon and diphoton final states depending on the dark fermion masses. 
The Feynman diagrams for diphoton final states are shown in Fig.~\ref{fig:diphoton}. 
The Feynman diagrams for digluon are similar to diphoton, except that only the dark tops are the particles running in the loop in the digluon decay. 
The tree-level decay partial width is
\bea
	\Gamma (s \to h^+ h^-) &=& \frac{g_{sh^+h^-}^2 }{8\pi m_s} \sqrt{1- \frac{4 m_{h^+}^2}{m_s^2}}, \\
	\Gamma (s \to {\mathcal N}{\mathcal N}) &=& \frac{y_N^2 m_s}{16\pi } \sqrt{1- \frac{4 m_{{\mathcal N}}^2}{m_s^2}}.
\eea
The loop-induced partial decay width in the digluon channel is 
\bea
\Gamma (s \to g g) 
&=& \frac{\alpha_s^2 m_s^3}{128 \pi^3} 
\left| \delta_F C[F]\frac{2g_{sff}}{m_f} A_{1/2}(\tau_f)   \right|^2,
\eea
where  $\tau_f\equiv4m_f^2/m_S^2$ and
\bea
&&A_{1/2}(\tau)=-2\tau(1+(1-\tau)f(\tau)), \\
&&f(\tau)=\left\{\begin{array}{ll}
\left(\sin^{-1}\sqrt{\frac{1}{\tau}}\right)^2, & \tau\geq1\\
-\frac{1}{4}\left(\log\frac{1+\sqrt{1-\tau}}{1-\sqrt{1-\tau}}-i\pi\right)^2, & \tau<1
\end{array}\right..
\eea
Here $\delta_i = 1/2$ for real field $i$ and 1 otherwise. The factor $1/2$ is to  account for lack of conjugate diagram for a real field.
$C[i]$ is the Dynkin index of the field representation, defined as ${\rm Tr}[T_i^a T_i^b] = C[i]\delta^{ab}$. 
For dark top $T$, we have $\delta_T = 1, C[T] = 1/2, g_{sff} = y_T$.
The diphoton partial width is
\bea
\Gamma (s \to \gamma \gamma) 
&=& \frac{\alpha^2 m_s^3}{1024 \pi^3} 
\left| N_{c,f} Q_f^2 \frac{2 g_{sff}}{m_f} A_{1/2}(\tau_f)  \right. \nn\\
&&\left.+ N_{c,S} Q_S^2  \frac{ g_{sSS}}{m_S^2} A_{0}(\tau_S) \right|^2,
\label{eq:diphotondecay}
\eea
where $A_{0}(\tau) = x - x^2 f(x)$, $N_{c,i}$ and $Q_i$ are color factor and electric charge of the particle $i$. 
For dark tops, we have $N_c = 3, Q_T = \frac23$ and for charged Higgs, we have $N_c = 1, Q_{h^+} = 1$.
The loop-induced $WW$ and $ZZ$ partial width is estimated to be 
\bea
	\Gamma_{\gamma\gamma} : \Gamma_{Z\gamma} : \Gamma_{Z Z} = 1: 2 \tan^2\theta_W : \tan^4\theta_W 
\simeq 1: 0.6 : 0.09,\nn\\
\eea
where $\theta_W$ is the weak mixing angle. Therefore, in following discussion, we neglect the branching ratio for loop-induced $WW$ and $ZZ$ final states.

\begin{figure}
  \includegraphics[width=0.22\textwidth]{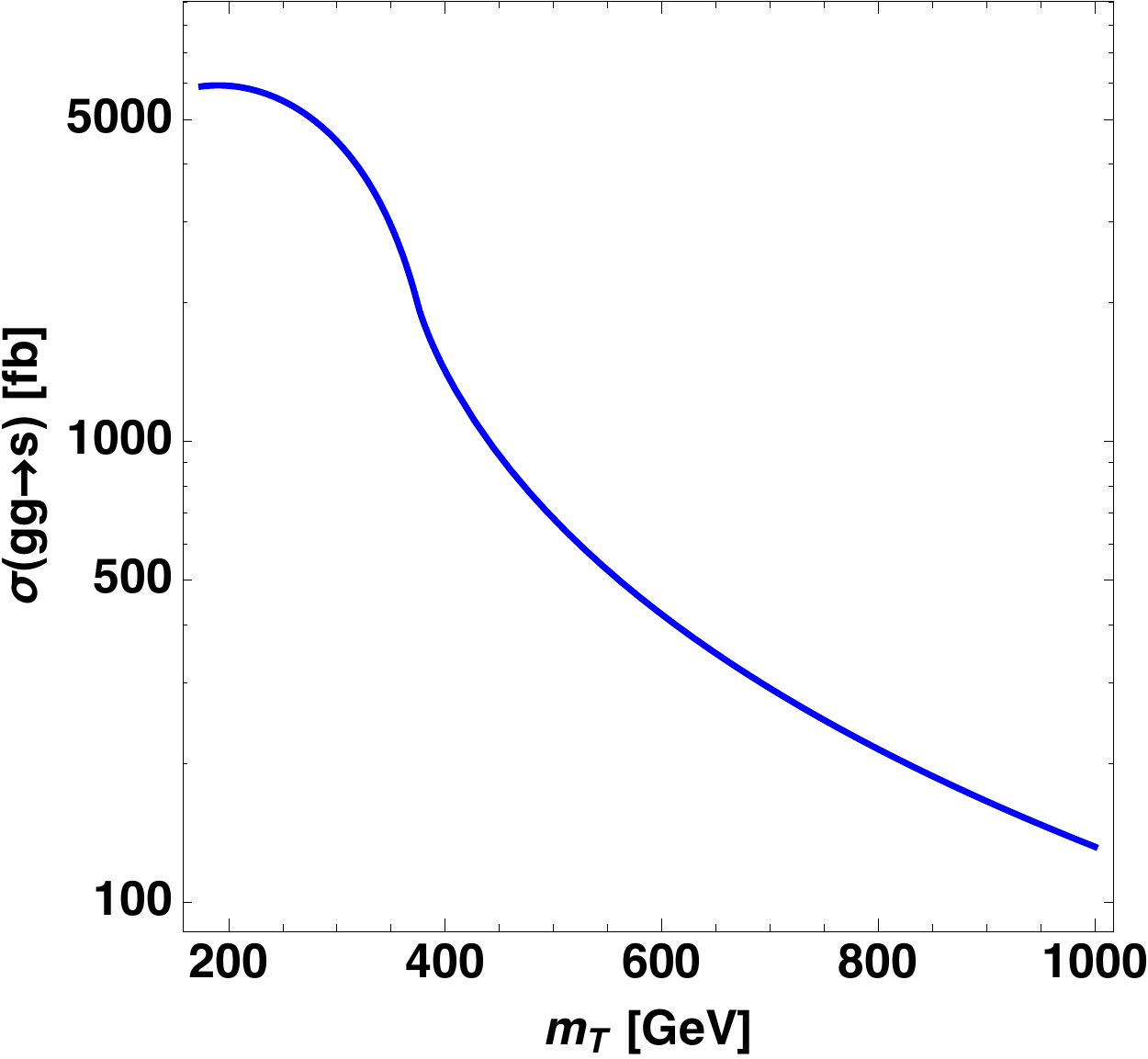} 
  \includegraphics[width=0.22\textwidth]{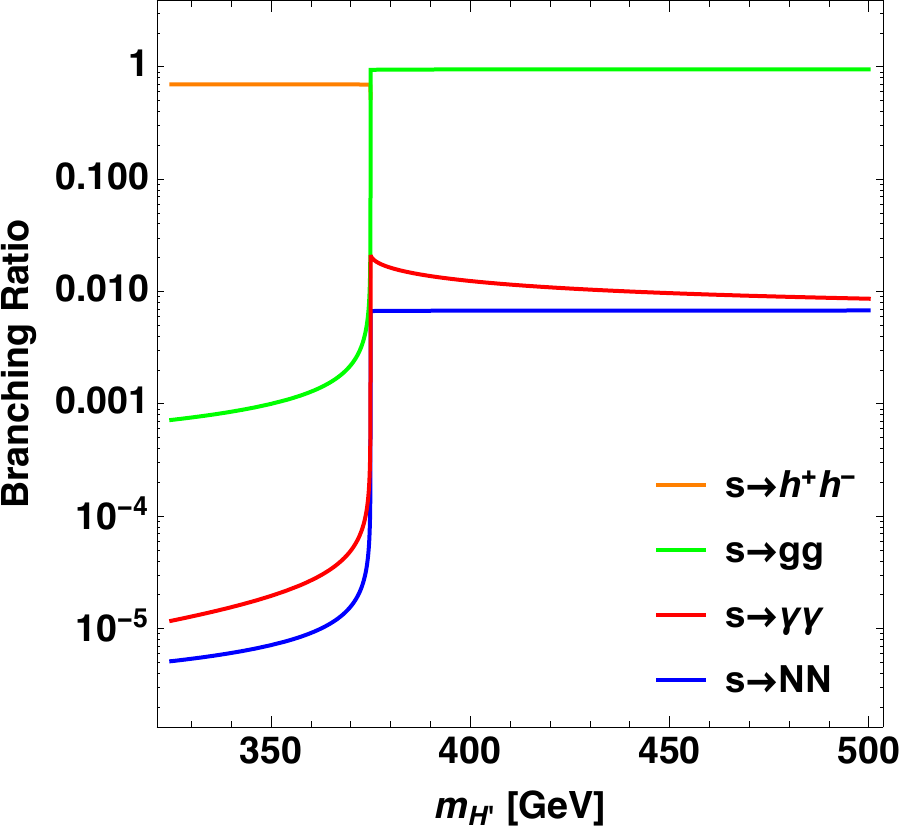} 
\caption{\label{fig:branch}  The production cross section of the 750 GeV scalar $s$ as the function of the $m_T$ (left panel), and its dominant decay branching ratios as the function of the $m_{h^\pm}$ (right panel). Here we assume $y_T = 1$, $y_{N'} = 0.005$, and $m_{T'} = m_T$, $m_{h'} = m_{h^\pm} = m_A$. In right panel, $m_T > m_s/2$ is also assumed. }
\end{figure}

Following discussions in sec.~\ref{sec:collider}, we assume that the masses of the dark tops are degenerate, and the masses of the inert Higgs components are also degenerate due to radiative neutrino mass generation and $S, T$ parameter constraints.
Fig.~\ref{fig:branch} shows the production cross section of the 750 GeV scalar $s$ as the function of the $T$ mass, and the decay branching ratios as the function of the charged Higgs mass. 
We calculate the production cross section of the 750 GeV $s$ using
\bea
	\sigma(gg \to s) = \frac{\Gamma_{s\to gg}}{\Gamma_{s\to gg}^{\rm SM}} \sigma^{\rm SM}(gg \to s),
\eea
where $\sigma^{\rm SM}(gg \to s)$ is taken to be its NNLO rate ($742$ fb) at 13 TeV LHC~\cite{Heinemeyer:2013tqa}.
%
%
From Fig.~\ref{fig:branch} (left panel) we see that the production cross section decreases dramatically as the dark tops $T(T')$ get heavier. 
Therefore, dark top masses need to be not so heavy: around $400 \sim 600$ GeV. 
On the other hand, for $m_T \sim 400 - 600$ GeV, the all hadronic top pair plus MET searches at the LHC put very strong constraints on the parameter space, as shown in Fig.~\ref{fig:heavyT} (lower panel). To avoid these constraints, the dark top needs to be close to the stealth region with mass
\bea
	m_{T} - m_{h^\pm} \lesssim 180 \, {\rm GeV}. 
\eea
On the diphoton decay branching ratio, we calculate the partial decay width using Eq.~\ref{eq:diphotondecay}. 
Here we assume that the dark tops are degenerate and heavier than $m_s/2$. 
The invisible decay $s \to {\mathcal N}{\mathcal N}$ is expected to be small, because the coupling $y_N$ is favored to be small to explain the radiative neutrino mass. 
From the Fig.~\ref{fig:branch} (right panel), we note that if $m_{h^\pm} < m_s/2$, the tree-level decay $s \to h^+ h^-$ dominates; while if $m_{h^\pm} > m_s/2$, the loop-induced diphoton and digluon decays will dominate.
We also note that the DM mass is irrelevant in the $s$ production and decay branching ratios by taking the invisible decay partial width is small. 
%
More interestingly, the diphoton decay branching ratio has an enhancement near the threshold $m_{h^\pm} \simeq m_s/2$. 
The reason is that  the functions $A_{1/2}(\tau)$ and $A_0(\tau)$ defined in Eq.~\ref{eq:diphotondecay} get large enhancements near the threshold region with $\tau \to 1$.
Therefore, if the masses of the dark tops or the inert Higgs are close to half of the $s$ mass, the diphoton
signal rate could be greatly enhanced.

\begin{figure}
  \includegraphics[width=0.36\textwidth]{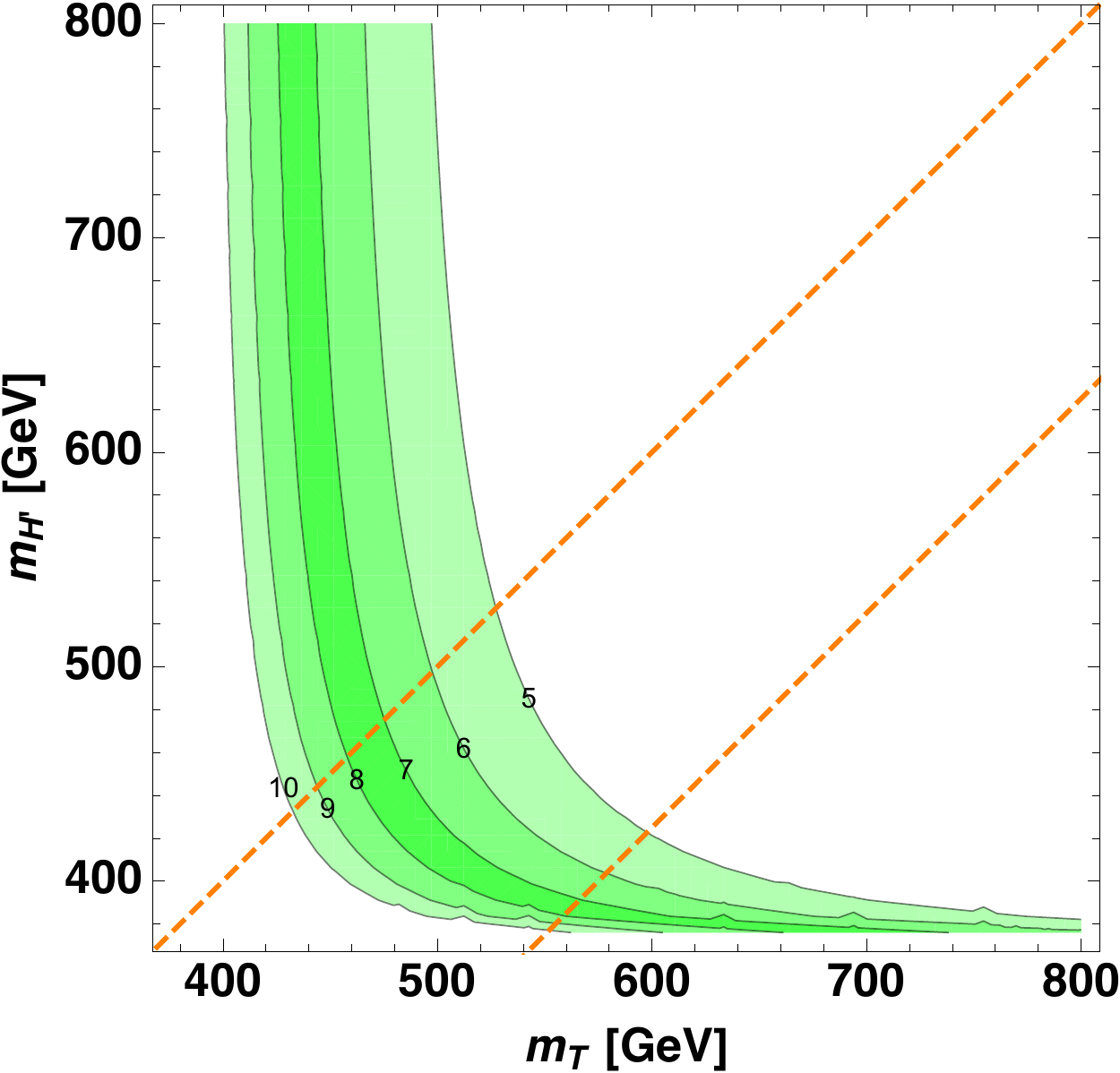}
\caption{\label{fig:signal}   The contours on production cross section times diphoton branching ratio in the $(m_{T}, m_{h'})$ plane. The green region shows the diphoton signal rate $5 \sim 10 $ fb. The contour labels show the diphoton rate in fb. Here we take $y_T = 1$, $y_{N'} = 0.005$, $\lambda_{h'} = 1.0$, $u = 1.2$ TeV, and $m_{T'} = m_T$, $m_{h'} = m_{h^\pm} = m_A$. 
The two dashed lines shows the stealth region in between. }
\end{figure}

\begin{figure}
  \includegraphics[width=0.36\textwidth]{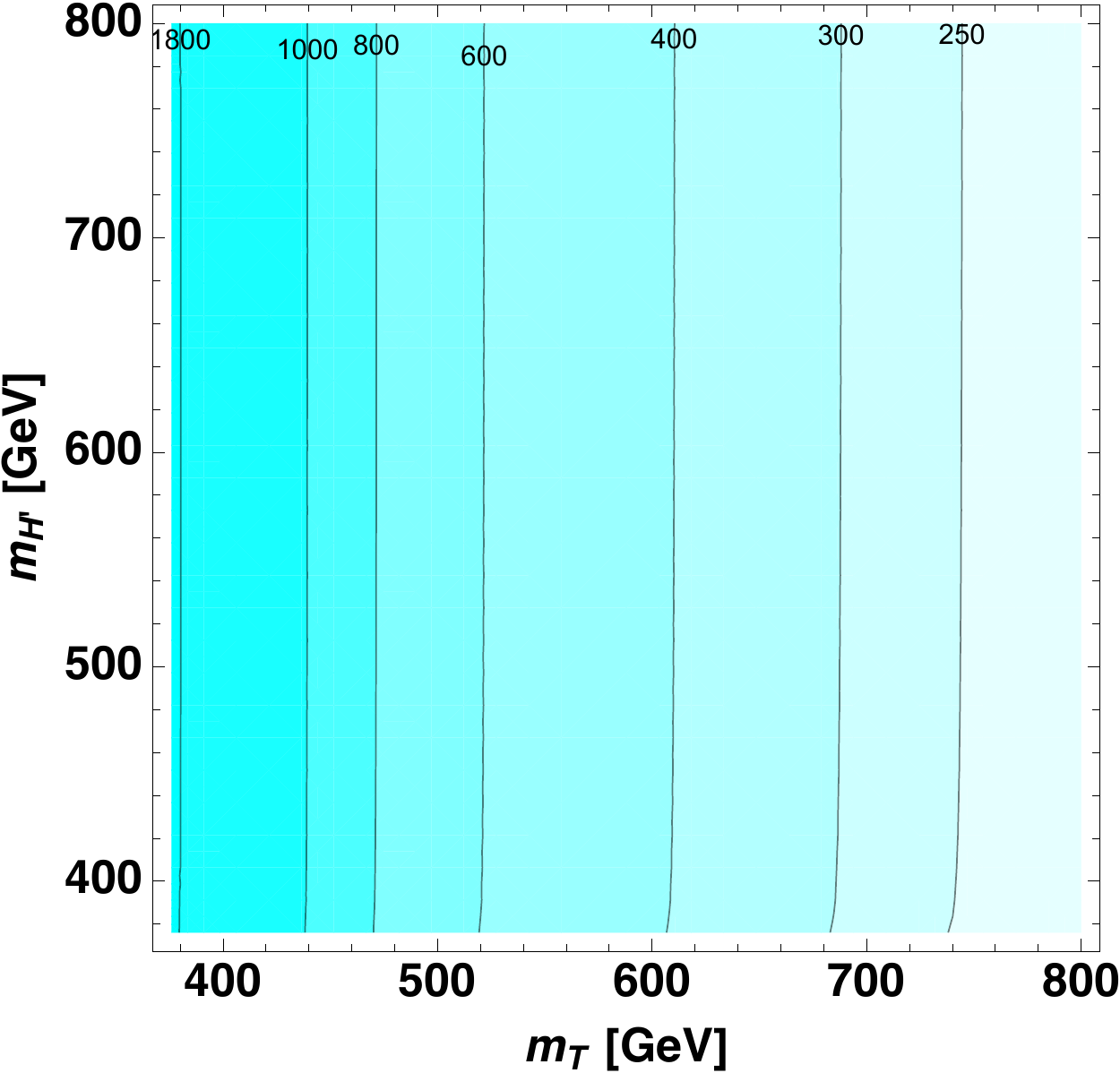}
\caption{\label{fig:gluon}   The contours on production cross section times digluon branching ratio in the $(m_{T}, m_{h'})$ plane. The contour labels show the digluon rate in fb. Here we take $y_T = 1$, $y_{N'} = 0.005$, $\lambda_{h'} = 1.0$, $u = 1.2$ TeV, and $m_{T'} = m_T$, $m_{h'} = m_{h^\pm} = m_A$. }
\end{figure}

We calculate the total diphoton rate in the favored parameter space: not-so-heavy $m_T \simeq m_{T'}> m_s/2$, $m_{h'} = m_{h^\pm} = m_A > m_s/2$, very small $y_{N'}$, small mixing angle $\sin\phi$. 
Given the typical values of the coupling strength $y_T$ and $\lambda_{h'}$, the total rate is 
around $5 \sim 10 $ fb. 
Fig..~\ref{fig:signal} shows the contours on production cross section times diphoton branching ratio in the $(m_{T}, m_{h'})$ plane. The green region shows the parameter region which obtains diphoton signal rate $5 \sim 10 $ fb. We find that the dark particles with several hundred GeVs could fit the diphoton excess well.
Specifically, the dark tops are around $450 \sim 550$ depending on the charged inert Higgs mass~\cite{Chao:2015ttq}.
If the charged Higgs mass is near $m_s/2$ threshold, the diphoton branching ratio will be greatly enhanced, and thus masses of the dark tops could be heavier, around $600 \sim 700$ GeV.
If the down-type dark fermions, such as $B$, $E$, are not decoupled, we expect there will be larger parameter space which could fit the diphoton excess.
Assumming $Z'$ is lighter than half of the $s$ mass, and more than one generations of the light dark fermions, it is also possible to obtain large decay width of the scalar $s$. 
The light dark fermions $T$, $B$, $N$, $E$ could greatly enhance the production cross section of the $s$, while the tree-level decay $s \to Z' Z'$ could contribute to the possible large width of the scalar $s$. 
%
%
%

We also need to validate this model through non-observation of this 750 GeV resonance in other channels, such as dijet, $WW$, $ZZ$, $tt$, and invisible decay final states.
The current limits on 750 GeV resonance are $\sigma\cdot {\rm Br}_{\rm dijet} < 1800 $ fb~\cite{CMS8diphoton}, $\sigma\cdot {\rm Br}_{ZZ} < 22 $ fb~\cite{Aad:2015kna}, $\sigma\cdot {\rm Br}_{WW} < 38 $ fb~\cite{Aad:2015agg}, $\sigma\cdot {\rm Br}_{tt} < 600 $ fb~\cite{CMS8tt}, $\sigma\cdot {\rm Br}_{\rm inv.} < 3000 $ fb~\cite{Aad:2015zva}. 
In favored parameter space with small $y_{N'}$ and small mixing $\sin\phi$, it is
obvious that the branching ratios of the $WW$, $ZZ$, $tt$, and invisible decay final states are small and 
under the current limits. 
On the other hand, the dijet channel is dominant in the favored parameter space. 
We need to check whether the dijet rate in the signal region is smaller than $1800$ fb at the 13 TeV LHC.
Fig.~\ref{fig:gluon} shows the contours of the digluon rates in the $(m_T, m_{h^\pm})$ plane, %
given the same parameters as the diphoton signature. 
The cyan contours in Fig.~\ref{fig:gluon} show that the dijet rate in the signal region is much smaller than current dijet limit.
%


\section{Conclusion}
\label{sec:conclude}

We investigated hidden gauged $U(1)$ models which unify the scotogenic neutrino ($\equiv$ neutrino portal dark matter) and flavor portal dark matter. 
In this scenario, not only the left-handed neutrino has its dark partner via Yukawa interaction, but also all other left-handed fermions in the standard model have its dark fermion partner. 
Similar to the standard model Higgs, the Yukawa couplings between the left-handed fermions and their dark fermion partners are through a single inert Higgs boson.
The neutrinos obtain their masses radiatively, the same as the scotogenic neutrino model.
Due to existence of the flavor portal interactions, the neutral components of the inert Higgs becomes a viable dark matter. 
More interestingly, this unified scenario provides us richer collider phenomenology than the scotogenic neutrino and flavor portal dark matter alone.

We focused on the case that the lightest dark neutrino is the dark matter candidate, and calculated its relic density and direct detection rate. 
Then we discussed collider signatures for the dark particles: the dark partners $T(T'), h^\pm (A, h')$ and the gauge boson $Z'$ and its associated scalar $s$.
According to the existing bounds from neutrino mass, LUX, LEP and LHC data, we obtained the favored parameter region: small $y_{N'}$, degenerate inert Higgs components $m_{h'} \simeq m_{A} \simeq m_{h^\pm}$, and $m_T, m_{h^\pm}$ in the stealth region, and the dark matter mass in a very broad range.  
Given the favored parameter region, we addressed the scalar $s$ could be the 750 GeV diphoton resonance reported by ATLAS and CMS.
We expect this model could be further tested and falsified by high luminosity LHC Run-2 data.


\begin{acknowledgments}
The author thanks Michael Ramsey-Musolf, Mark Wise, and Clifford Cheung for useful discussion and comments during the visit of Caltech. This work was supported by DOE Grant DE-SC0011095. 
\end{acknowledgments}



\end{document}